\documentclass{article}
\usepackage{hyperref,amssymb,amsmath,mathrsfs,bm,graphicx}
\usepackage{amsmath,amssymb,amsfonts,amsthm,dsfont,dcolumn}
\usepackage{graphicx,wrapfig}

\usepackage{epsfig,graphics}
\usepackage{color}
\begin{document}
\centerline{\large\bf FIXED MESH REFINEMENT IN THE CHARACTERISTIC }
\centerline{\large\bf FORMULATION OF GENERAL RELATIVITY}
\vspace*{0.245truein}
\centerline{\footnotesize{\large W. Barreto\footnote{Centro de F\'\i sica Fundamental,
Facultad de Ciencias, Universidad de Los Andes, M\'erida, Venezuela}$^{,2}$, H. P. de Oliveira\footnote{Departamento de F\'\i sica Te\'orica, Instituto de F\'\i sica A. D. Tavares, Universidade do Estado do Rio de Janeiro,  R. S\~ao Francisco Xavier, 524, Rio de Janeiro 20550-013, RJ, Brasil}, B. Rodriguez-Mueller\footnote{Computational Science Research Center, San Diego State University, United States of America}}}
\baselineskip=12pt
\vspace*{0.21truein}
\date{\today}

\begin{abstract}
We implement a spatially fixed mesh refinement under spherical symmetry for the characteristic formulation of General Relativity. The Courant-Friedrich-Levy (CFL) condition lets us deploy an adaptive resolution in (retarded-like) time, even for the nonlinear regime. As test cases, we replicate the main features of the gravitational critical behavior and the spacetime structure at null infinity using the Bondi mass and the News function. Additionally, we obtain the global energy conservation for an extreme situation, i.e. in the threshold of the black hole formation. In principle, the calibrated code can be used in conjunction with an ADM 3+1 code to confirm the critical behavior recently reported in the gravitational collapse of a massless scalar field in an asymptotic anti-de Sitter spacetime. For the scenarios studied, the fixed mesh refinement offers improved runtime and results comparable to code without mesh refinement. 

\vspace*{0.21truein}
\noindent Key words: Numerical Relativity; Characteristic Formulation; Fixed Mesh Refinement; Critical Behaviour; Asymptotically AdS Spacetimes; AdS/CFT correspondence; Holographic Principle.
\end{abstract}
\section{Introduction}


{The gravitational critical behavior as originally discovered by Choptuik \cite{c93} is well understood and seems to be an ubiquitous phenomenon. It emerges in many contexts,  including when the gravitational collapse of a massless scalar field takes place in asymptotically anti de Sitter (AdS) spacetimes. But there are new features recently reported by Santos-Oliv\'an and So\-puerta \cite{ss16}, \cite{ss16b}: A series of critical points arises, branching with and without mass gap. Computationally, the calculation of the gravitational critical collapse is challenging, especially so for multiple 'cascading' critical points. There are two ways to attack these issues: i) Using Adaptive Mesh Refinement (AMR) \cite{bo84}, \cite{c89}, \cite{hs96}, \cite{pl04}; ii) Following the null geodesics \cite{g95}, \cite{petal05}. Here we report an implementation and testing of the Fixed Mesh Refinement (FMR) approach in the characteristic formulation of Numerical General Relativity. The final purpose is to use the developed code in combination with other code, which employs Domain Decomposition and the Galerkin-Collocation method in the ADM 3+1 formulation \cite{dopr13}. With this Characteristic-Cauchy merging we expect to make a future independent confirmation of the multiple critical points and as well as uncover fine-grained structure in asymptotically AdS spacetimes.}


{A massless scalar field, in the strong field limit near the formation of a black hole, under spherical symmetry and minimally coupled to gravity, displays: (i) a critical behavior of type II with a very small black hole mass; (ii) an unstable naked singularity by fine-tuning generic initial data; (iii) a power law mass scaling and shows discrete self-similarity. Critical behavior of type I is found when a massive scalar field (a Compton wavelength) is considered \cite{bcg97}, \cite{ss91}. For a review on the critical phenomena for gravitational collapse, including quantum extensions, see Ref. \cite{gm07}.
Recently we have been involved in calculations of the gravitational critical behavior with mass gap, evolving a scalar field kink \cite{bcdrr16}. It is interesting enough to explore and confirm the new features  \cite{ss16b} of the gravitational collapse of a massless scalar field in asymptotically AdS spacetimes. Near the multiple critical points, the mass spectra show a power law with and without mass gap. What is the meaning of this in the AdS/CFT dictionary?}


{The AdS spacetime is the (maximally symmetric) solution for the vacuum Einstein equations with a negative cosmological constant. The boundary of AdS spacetime plays a fundamental role in the Holographic Principle implementation. The Holographic Principle establishes the equivalence between two Universes with different dimensions obeying different physical laws \cite{bekenstein}. Maldacena \cite{maldacena} reported one mathematical realization of this principle: one 5-dimensional (5-D) spacetime corresponding to a hologram at the boundary of a 4-dimensional (4-D) spacetime. 
A black hole in a 5-D spacetime is equivalent to thermal radiation in the 4-D Hologram; they have the same entropy, but the physical origin is different for each case \cite{thoof}-\cite{susskind}. No experiment can establish the difference between these two descriptions of the Universe.  
Thus, understanding gravitational collapse for asymptotically AdS spacetimes is important. The simplest model for this scenario is a massless scalar field minimally coupled to gravitation, the Einstein-Klein-Gordon (EKG) system.}

{The EKG system has been useful in gravitational collapse and black hole dynamics to develop a better understanding of: discovery of the gravitational collapse critical behavior \cite{c93}, modeling and simulation of a black hole binary system \cite{pretorius}-\cite{Bakeretal}, simulations to analyze the gravitational radiation detected by LIGO \cite{LIGO}.}
{The instability problem for EKG collapse in an asymptotically AdS spacetime remains open. The EKG system plays a fundamental role as a toy model to discover new phenomena,  translating them to observational Physics across the Holographic  Principle 
\cite{lehner}.}

{Mesh refinement can be static or dynamic, that is, fixed or adaptive. When an (uniform) unigrid setting is not enough to resolve the fine structure, each problem determines the type of multigrid. For instance, if a discontinuity -typically formed with a shock wave- is moving, then the AMR is the right answer; if confined to some region, then FMR should be enough. In any case, if we know in advance how to handle regions with different resolutions, then FMR paves the way for AMR \cite{shh04}. Here we adopt the practical point of view that a FMR code is sufficient in its own right for this problem because we know the location of the problem spot in space.

{We report a characteristic code with FMR intended to be combined with both Domain Decomposition and Galerkin-Collocation. The goal of this work is to create a FMR code which preludes and enables accurate and efficient study of critical behavior for the EKG system in an asymptotic AdS spacetime \cite{ss16}, \cite{ss16b}.} {As far as we know, the FMR method has not been implemented in the characteristic formulation of Numerical General Relativity. Basically it consists in a recursively static domain decomposition, which requires a previous knowledge of the problem. In our case we implement the FMR only in one coordinate (radial and null) which, given the structure $R\times S^2$, can be easily extended to higher dimensionality.}


{We organize the work as follows. In section 2, we write the field equations (including the cosmological constant) for the Bondi-Sachs coordinates under spherical symmetry. We briefly explain how the deal with the origin and with infinity, mentioning the numerical methods employed to solve the equations. In section 3 we revisit the well established critical behavior (without cosmological constant) as discovered by Choptuik. In section 4 we show the results and tests of the implemented FMR in the characteristic formulation. Finally, we discuss our results and conclude in section 5.}

\section{Setup}

\subsection{The field equations}
We use the Bondi-Sachs metric \cite{bvm62}, \cite{s62} under spherical symmetry \cite{b64} 
\begin{equation}
ds^2=e^{2\beta}du[(V/r)du+2dr]-r^2(d\theta^2+\sin^2\theta d\phi^2), \label{el}
\end{equation}
where $\beta$ and $V$ are functions of $u$ and $r$. Here $u$ is a timelike coordinate; in a flat spacetime $u$ is just the retarded time. Therefore, surfaces $u=$constant represent null cones open to the future; $r$ is a null coordinate ($g_{rr}=0$) such that surfaces $r=$constant are spheres; $\theta$ and $\phi$ are the usual angular coordinates.
Thus, we write the Einstein-Klein-Gordon system, including the cosmological term, as:
\begin{equation}
\beta_{,r}=2\pi r \Phi_{,r}^2,
\label{ekg_a}
\end{equation}
\begin{equation}
V_{,r}=e^{2\beta}(1-3r^2/\ell^2),
\label{ekg_b}
\end{equation}
\begin{equation}
2(r\Phi)_{,ur}=r^{-1}(rV\Phi_{,r})_{,r},
\label{ekg_c}
\end{equation}
where the comma represents a partial derivative with respect to that coordinate, 
$\Phi=\Phi(u,r)$ is a massless scalar field and $\ell$ is the AdS length scale, which is related to the cosmological constant $\Lambda$ by $\ell^2=-3 /\Lambda$. 
This is an initial-boundary problem. Specifying the initial null data $\Phi(u_0,r)$ at the initial time $u_0$, and using the gauge freedom $\Phi \rightarrow \Phi\,\, +$ constant, we set $\Phi(u_0,\infty)=0$ to solve the problem. We assume that $\Phi(u_0,r)$ is not singular at $r=0$. We make $\Lambda=0$ for the purposes of this paper, then the spacetime described by metric (\ref{el}) is asymptotically flat. Note that for a non-zero cosmological constant we need to transfer the evolved initial data from an interior (characteristic) to an exterior (Cauchy) asymptotically AdS. Otherwise, we have to use the affine metric of Chesler and Yaffe \cite{cy14}, which let us reach asymptotically the AdS spacetime using characteristics. 

The resulting metric does not take an asymptotic Minkowski form in the limit $r\rightarrow\infty$ of future-null-infinity (${\mathcal J}^+$). Because $\beta(u,\infty)=H(u)$, the Bondi time $u_B$ for a Minkowski frame at ${\mathcal J}^+$ relates to the proper time $u$ along the central geodesic by means of
\begin{equation}
\frac{du_B}{du}=e^{2H}.\label{rs}
\end{equation}
The coordinates $(u_B,r,\theta,\phi)$ constitute a standard Bondi frame whose line element is given by (\ref{el}) with the replacements $V\rightarrow V_B=e^{-2H}V$ and $\beta\rightarrow \beta_B=\beta-H$. Bondi time is more convenient to explore asymptotic quantities such as the mass and news function. The central time is more convenient to deal with horizons. A horizon forms at a finite central time $u^{\mathcal H}$ but at an infinite Bondi time $u_{B}^{\mathcal H}$, with a central redshift given by (\ref{rs}).

\subsection{Near infinity}
At ${\mathcal{J}}^+$: $g(u_0,\infty)=Q(u_0)$ and $\partial_r^n g(u_0,\infty)=0$, for
$n\ne 0$, where $g=r\Phi$ and $Q(u)$ is the scalar monopole moment.
Assuming the scalar field has an asymptotic expansion 
\begin{equation}
\Phi(u,r)=\frac{Q(u)}{r}+ \frac{c_{NP}}{r^2} + O(r^{-3}),
\end{equation}
the hypersurface equations (\ref{ekg_a}) y (\ref{ekg_b}) lead to
\begin{equation}
\beta(u,r)=H(u)-\frac{\pi Q^2(u)}{r^2}+ O(r^{-3}),
\end{equation}
\begin{equation}
V(u,r)=e^{2H}\left( r-2M(u)+\frac{\pi Q^2(u)}{r}\right) + O(r^{-3}),
\end{equation}
where $H(u)$ and $M(u)$ are integration functiones with physical interpretation, as we shall see. The expansion of the wave equation (\ref{ekg_c}) implies ${c_{NP}}_{,u}=0$, where $c_{NP}$ is the Newman-Penrose constant for the scalar field.
On physical grounds the Bondi mass can be defined as
\begin{equation}
M(u)=\frac{1}{2}e^{-2H}r^2(V/r)_{,r}|_{r=\infty},
\end{equation}
for which exists a mass loss equation given by
\begin{equation}
e^{-2H}\frac{dM}{du}=-4\pi N^2,
\end{equation}
where 
\begin{equation}
N(u)=e^{-2H}\frac{dQ(u)}{du},
\end{equation}
is the News function.
It can be shown that the Bondi mass and the scalar News can be written as \cite{gw92}, \cite{b14}:
\begin{equation}
M=2\pi\int^\infty_0 rVe^{2\beta}\Phi^2_{,r}dr,
\end{equation}
and
\begin{equation}
N=\frac{1}{2}e^{-2H}\int^\infty_0 \frac{V}{r}\Phi_{,r}dr.
\end{equation}


\subsection{Near the center} 
Near $r=0$ we adopt the conditions
\begin{equation}
\beta(u,r)=O(r^2) \,\, \mbox{;} \,\, V(u,r)=r+O(r^3), \label{ro}
\end{equation}
such that the metric reduces to the Minkowski form (polar null) along the central world line:
\begin{equation}
ds^2=du^2+2dudr -r^2(d\theta^2+\sin^2\theta d\phi^2).
\end{equation}

In the general dynamical case the scalar field is free at the center. For that
reason we make the expansion around $r=0$
\begin{equation}
\Phi(u,r)=\Phi_0(u)+r\Phi_1(u)+r^2\Phi_2(u).
\end{equation} 
In this case we get from (\ref{ekg_a})-(\ref{ekg_c})
\begin{equation}
V=r-\frac{2\pi}{3}\Phi_1^2 r^3+\frac{4\pi}{3}\Phi_1\Phi_2 r^4 + \mathcal{O}(r^5),
\end{equation}
\begin{equation}
\beta=\pi\Phi_1^2 r^2 + \frac{8\pi}{3}\Phi_1\Phi_2 r^3 + \mathcal{O}(r^4),
\end{equation}
\begin{eqnarray}
\dot\Phi_0&=&\Phi_1, \label{dotphi0}\\
\dot\Phi_1&=&\frac{3}{2}\Phi_2, \label{dotphi1}\\
\dot\Phi_2&=&\frac{4\pi}{9}\Phi_1^3-2\Phi_0^2\Phi_1. \label{dotphi2}
\end{eqnarray}
where overdot indicates a derivative with respect to $u$.

The scalar curvature 
\begin{equation}
R=8\pi e^{-2\beta}\Phi_{,r}\left(2\Phi_{,u}-\frac{V}{r}\Phi_{,r}\right).
\end{equation}
at the center is:
\begin{equation}
R(u,r=0)=8\pi \Phi_1^2.
\end{equation}

\subsection{Numerical methods}

{To solve the field equations we use a null cone evolution algorithm for nonlinear scalar waves developed in Refs. \cite{gwi92}, \cite{gw92} (the 1D Pitt code) adapted to the present setting as reported in \cite{b14}. The characteristic formulation in Numerical General Relativity is well documented by Winicour in Ref. \cite{winicour}. The algorithm is based upon the compactified radial coordinate $x=r/(R+r)$, so that ${\mathcal J}^+$ is represented by a finite grid boundary, with $x=0$ at the center and $x=1$ at ${\mathcal J}^+$. The code has been tested to be globally second order accurate. This code has been used to get global energy conservation near the critical behavior \cite{b14}. Additionally, we implement a four-level FMR, as explained in section 4, and a fourth order Runge-Kutta to solve the system of equations given by (\ref{dotphi0})-(\ref{dotphi2}).}

\section{Choptuik's solution}
Let $\mathcal{S}$ denote a solution of (\ref{ekg_a})-(\ref{ekg_c}). Choptuik \cite{c93} focused his attention on the family of one-parameter solutions $\mathcal{S}[p]$. Each solution is generated by evolving an initially incoming massless scalar field. Each family has the property that the parameter $p$ characterizes the strength of the self-interacting scalar field. There is a parameter value $p_{weak}$ such that in the limit $p\rightarrow p_{weak}$ the spacetime is flat. At the other extreme, there is a parameter value $p_{strong}$ such that as $p\rightarrow p_{strong}$ the end state of the evolution is a black hole. Between these two extremes, a critical value $p^*$ exists where black hole does not form nor the solution disperses. Assuming that $p_{weak} < p^* < p_{strong}$, solutions $\mathcal{S}[p^* < p_{weak}]$ and $\mathcal{S}[p^* > p_{strong}]$ are subcritical and supercritical, respectively.

Choptuik discovered (using AMR) that the spacetime is discretely self-similar when the initial data is fine-tuned to the critical parameter $p^*$. Spacetime is discretely self-similar (DSS) if it admits a discrete diffeomorphism $D_\Delta$ which leaves the metric $g$ invariant by a scale factor
\begin{equation}
(D^*_\Delta)^n g=e^{2n\Delta} g,
\end{equation}
where $\Delta$ is a dimensionless real constant and $n\in {\mathcal N}$.

Let us define
\begin{equation}
\tau=-\ln\big\{\frac{u^*-u}{u^*}\big\} \label{tau}
\end{equation}
and 
\begin{equation}
\rho=\frac{r}{u^*-u}=\frac{r}{u^*}e^\tau,
\end{equation}
where $u^*$ is a real number, which we call the accumulation time in the discrete
self-similar spacetime. 

The critical function $\zeta^*$ (scalar field or metric) satisfies the scaling relation 
\begin{equation}
\zeta^*(\tau,\rho)=\zeta^*(\tau-n\Delta,\rho-n\Delta), 
\end{equation}
and does not depend on the family of initial data, which means it is universal, that is, with a numerical value of $\Delta\approx 3.4$ for any initial data. But it is important to say that it occurs in a neighborhood $r=0$, at least as Choptuik reported. What does this behavior mean? If we freeze the critical evolution at some time, examine the profile $\zeta$ in some delimited region, and continue evolving for a $\delta u$ and reexamine the solution on a scale $e^\Delta$ times smaller than previously, we will see the same field (metric) profiles. If we then wait for an additional time interval $\delta u/e^\Delta$ and ``zoom in" by another factor of $e^\Delta$, we will see again the same profiles. Thus, a precisely critical configuration will be characterized by an infinite series of ``echoes" in the field patterns (as well as another form-invariant quantity) which arise from dynamics unfolding on increasingly smaller spatiotemporal scales.
For each family of initial conditions, we can find $p^*$ (to the limit of machine precision) using a binary search predicated on whether or not a black hole forms.
Another universality feature is the agreement of the profiles at late times, regardless of the initial pulse shape.
To generate universality and echoing we use subcritical initial data. 

The critical regime may also be studied using supercritical evolutions characterized by the formation of black holes. The black hole mass obeys the power law
\begin{equation}
M_{BH}\simeq c_f|p-p^*|^\gamma \label{pl}
\end{equation}
where $c_f$ are a family dependent constants, but $\gamma$ is a universal scaling exponent which has a numerical value of $\gamma\approx 0.37$. $\gamma$ is the same for a family of initial data. 
These results suggest that the black hole mass turns out to be infinitesimal in this model problem. Any detail that might appear in the specification of the initial data is washed out by the interaction between the scalar and gravitational fields. 

{The numerical values for $\Delta$ and $\gamma$ depend on the matter fields and the geometry, but under the present context, i.e., the EKG system without cosmological constant, the values of $\Delta$ and $\gamma$ obtained by numerical experimentation are the same for any initial data. $\Delta$ is a measure of the discrete self-similarity (echoing) and $\gamma$ is related with the largest Lyapunov exponent \cite{kha95}. They  characterize subcritical and supercritical behavior, respectively. A combination of them $\Delta/2\gamma$ explain the oscillations for the fine structure in the power law \cite{petal05}. The gravitational critical behavior has been found in many systems and is well documented in Ref. \cite{gm07} (and references therein). In the present context we use these well established results to test our extended code with FMR.} 
\section{Characteristic FMR}

The implementation of the FMR in the characteristic 1D Pitt code is very simple, and can be resumed in two steps:
\begin{itemize}
\item {\sc Fixed radial refinement}: up to the selected level, in the radial compact coordinate $x$;
\item {\sc Adaptive Time Step}: The spatial refinement determines the minimal CFL time step, which is adapted in the evolution up to the critical behavior.
\end{itemize}
\begin{figure}
\rotatebox{0}{\includegraphics*[scale=0.46]{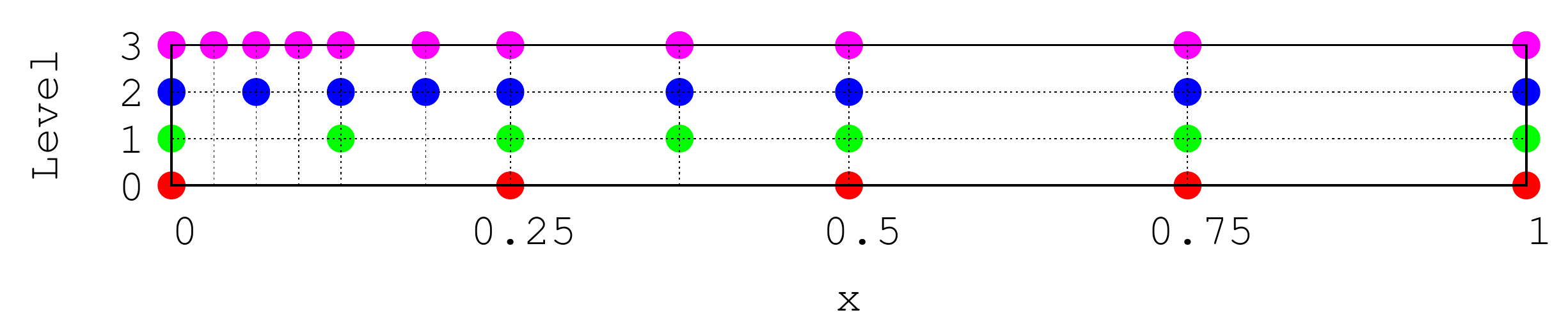}}  
\caption{One strategy for the FMR.}
\end{figure}  
We choose each refinement just on the middle of each (sub)grid to the left, with the same number of original points.
For instance, as indicated in Figure 1, if $N_x^{(n)}$ represents the number of grid points at the refinement Level $n$, we have $N_x^{(0)}=5$, $N_x^{(1)}=7$, $N_x^{(2)}=9$ and $N_x^{(3)}=11$; and grid sizes $\Delta x^{(0)}=1/(N_x^{(0)}-1)$, $\Delta x^{(n+1)}=\Delta x^{(n)}/2$, with $n=1,\,2,\,3$. Therefore the grid sizes are related in a proportion $8:4:2:1$. We can choose any other proportion or more levels to go with refinement. The selection of the refinement first point can be any interior point of the original (Level 0) grid and the refinement itself can be realized to the left ($x\rightarrow 0$) or to the right ($x\rightarrow 1$).

Now, knowing Choptuik's solution we replicate the echoing and power law. This can be considered as demanding tests of our FMR implementation. 
Using as initial condition
\begin{equation}
\Phi(0,r)=\lambda r^2 e^{-(r-r_0)^2/\sigma^2},
\end{equation}
\begin{figure}
\rotatebox{0}{\includegraphics*[scale=0.4]{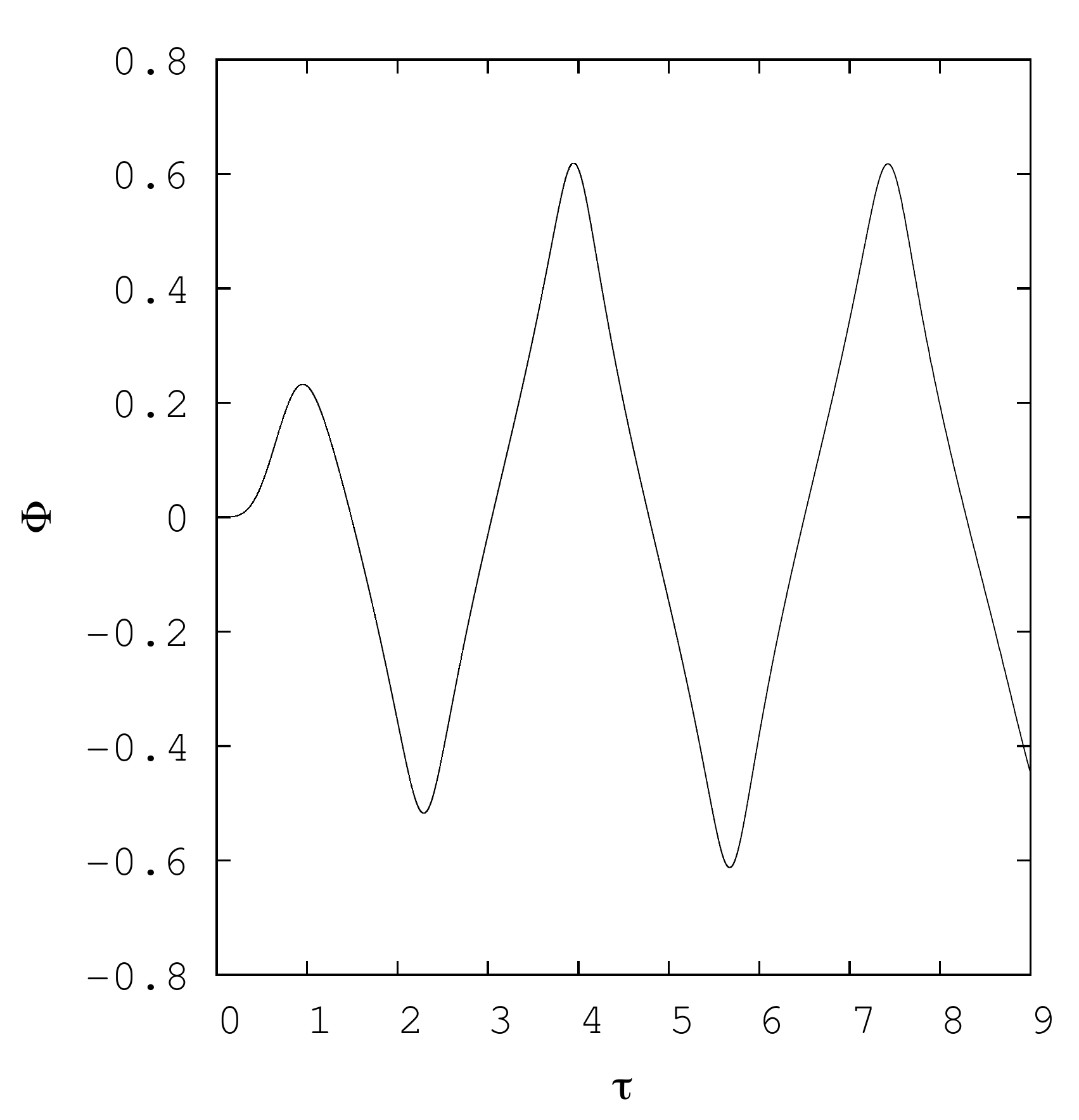}}  
\caption{Scalar field at $r=0$ as a function of $\tau$. It is apparent the periodicity ( one and a half cycle) in time with $\Delta\approx 3.4$. This calculation took 42 minutes on a 2.4 GHz Intel Core i5.}
\end{figure}
\begin{figure}
\rotatebox{0}{\includegraphics*[scale=0.4]{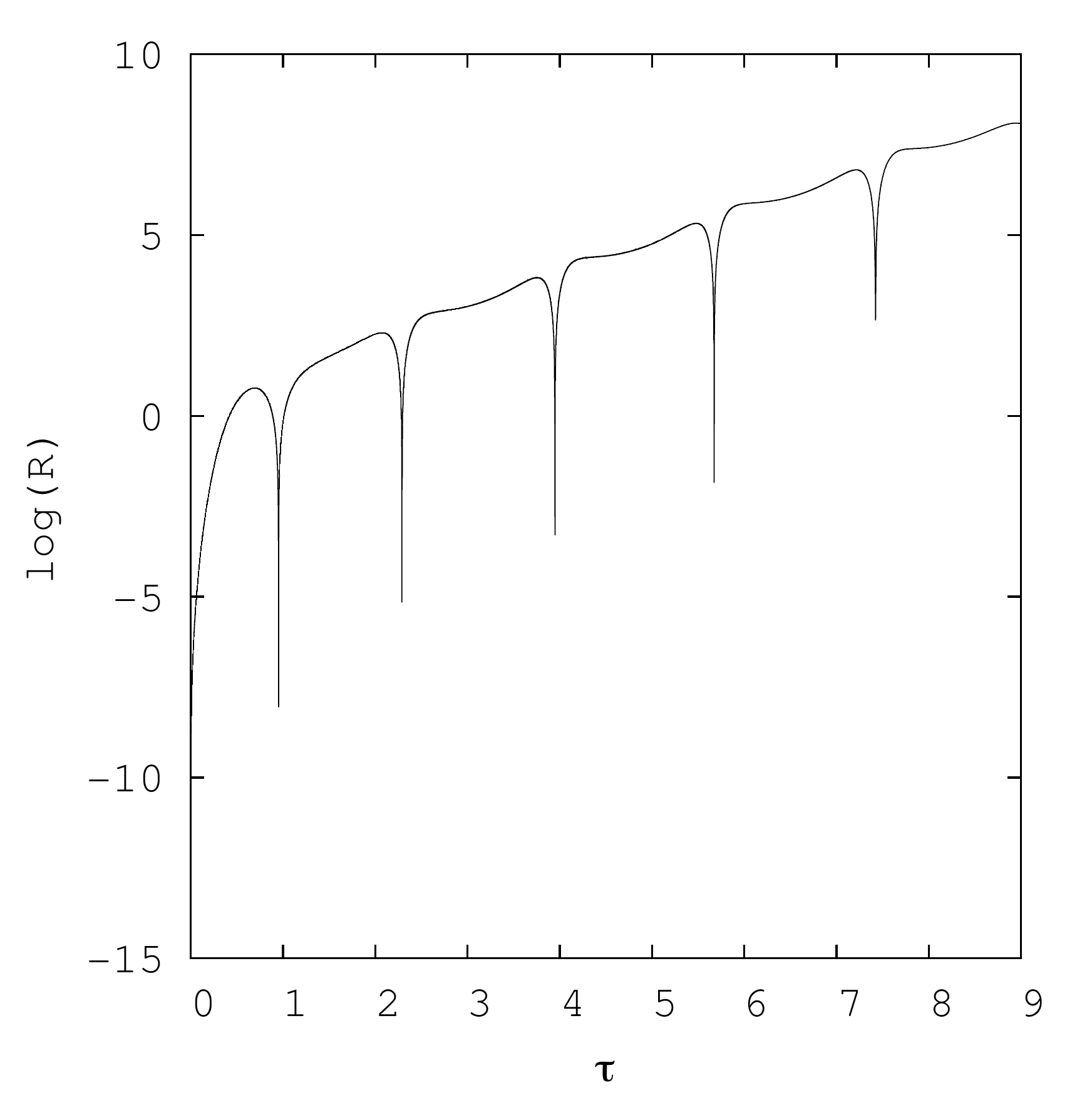}}  
\caption{Scalar curvature at $r=0$ as a function of $\tau$. It is apparent the periodicity  $\Delta/2$ in time (in three cycles).}
\end{figure}
with $r_0=0.7$, $\sigma=0.3$ and $\lambda=0.144930560446$, we show 
in Figure 2 the scalar field at $r=0$ as a function of $\tau$, using a grid refined from $N_x^{(0)}=10,001$ using the strategy shown in Fig. 1. 
That is, we increase in $10,001$ the number of points to the left of the midpoint in the previous Level to get $N_x^{(1)}=15,001$. In this specific case from Level $0$ to Level $3$ we get $N_x^{(3)}=25,001$. Our estimate accumulation time under the aforementioned conditions is $u^*=2.119620369$. Figure 3 displays the scalar curvature at $r=0$ as a function of $\tau$ \cite{hs96} and the same conditions of Fig. 2. The accumulation time is determined when the scalar curvature begins to decay (for the subcritical evolution closest to $\lambda^*$). 

\begin{figure}
\rotatebox{0}{\includegraphics*[scale=0.6]{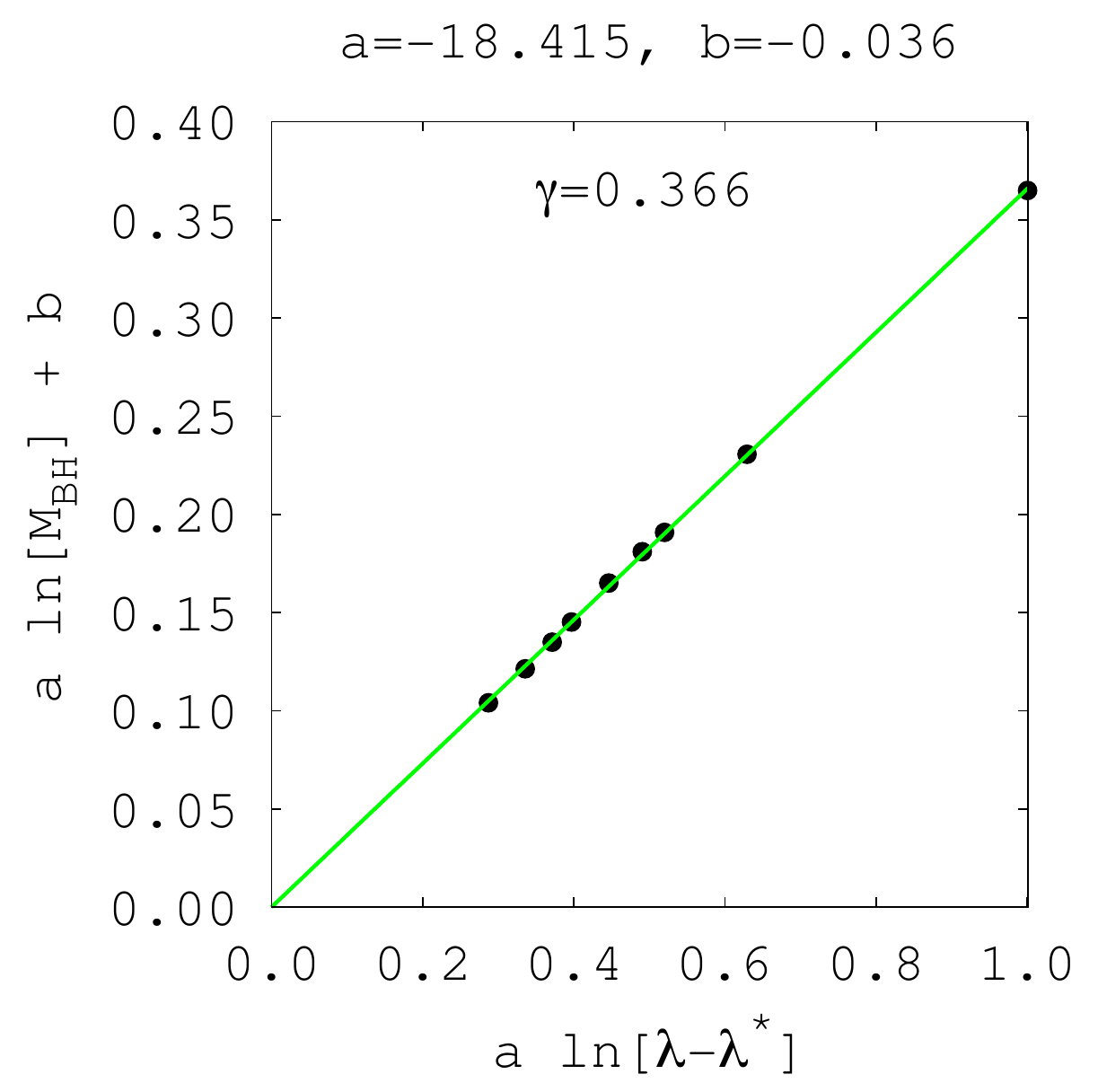}}  
\caption{Mass spectrum for the supercritical case $\lambda > \lambda^*$; the slope is $\gamma=0.366$.}
\end{figure}
\begin{figure}
\rotatebox{0}{\includegraphics*[scale=0.4]{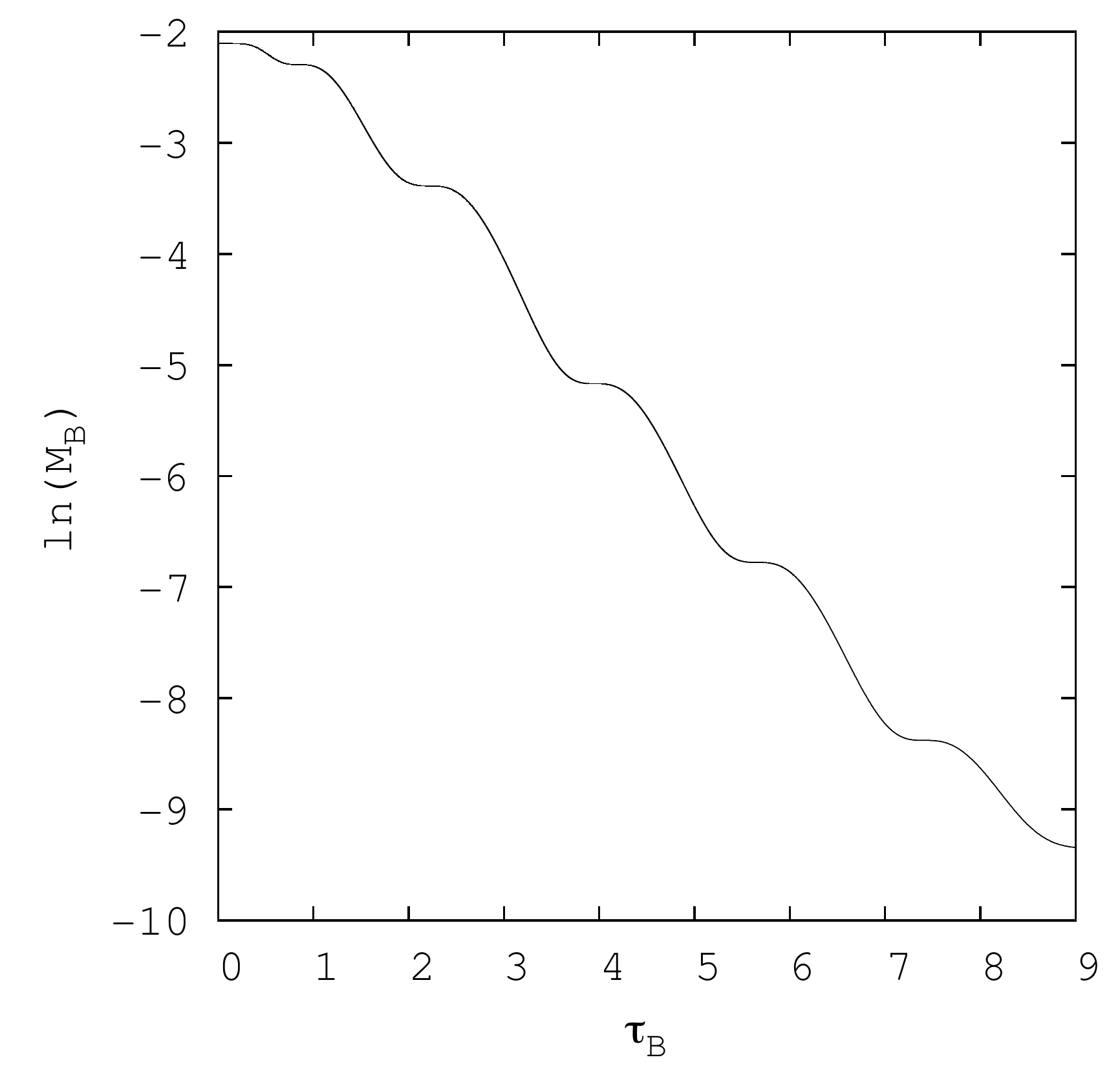}}  
\caption{Natural log of the Bondi mass as a function of $\tau_B$; the apparent periodicity is $\Delta/2$.}
\end{figure}
\begin{figure}
\rotatebox{0}{\includegraphics*[scale=0.4]{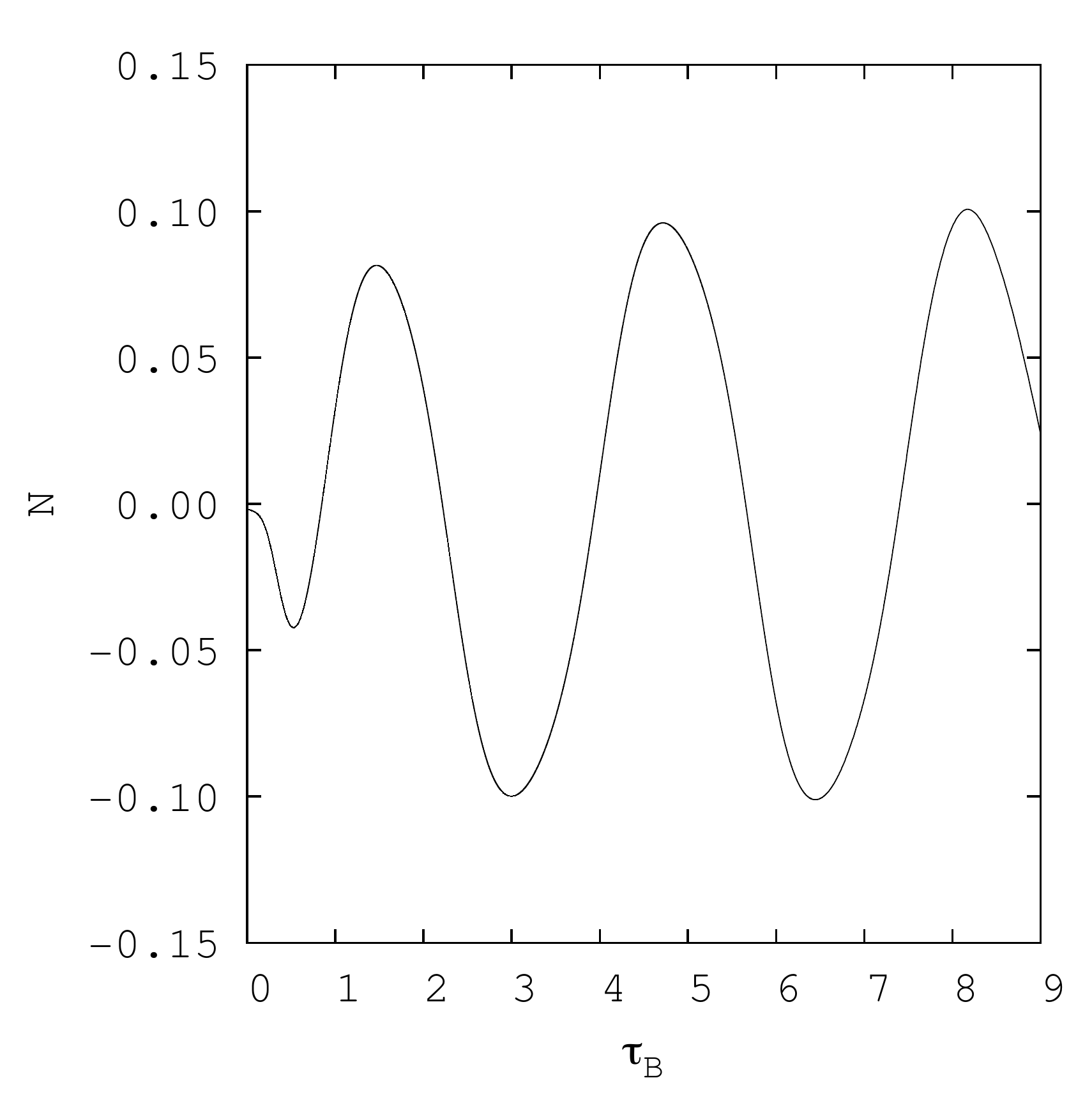}}  
\caption{News as a function of $\tau_B$; the apparent periodicity is $\Delta$.}
\end{figure}
\begin{figure}
\rotatebox{0}{\includegraphics*[scale=0.4]{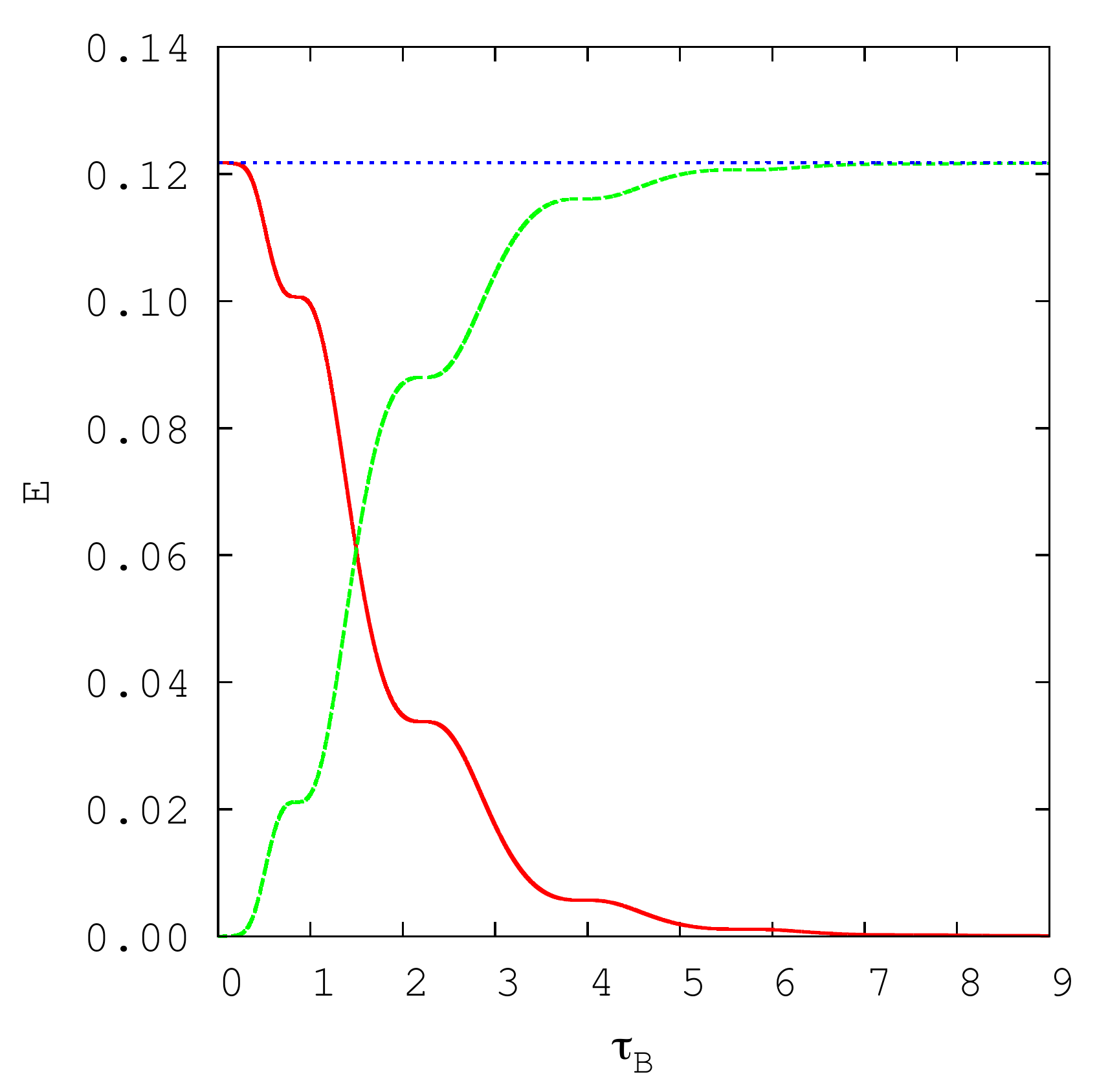}}  
\caption{Global energy conservation near critical behavior.}
\end{figure}

{Only for the sake of completeness we also replicate the power law for supercritical evolutions as shown in Fig. 4. We have used the same initial condition as in Fig. 1, with $N_x=10,001$ and $\lambda > \lambda^*$. We choose
$a$ and $b$ to normalize the abscissa and show the smallest black hole, following Choptuik \cite{c93}. The scaling exponent was fitted using a mean squared quadrature, resulting in $\gamma=0.366 \,(1\%)$. For this calculation we do not require the FMR code; each point takes 3 minutes. For each evolution, we picked up the minimum Bondi mass just before the black hole formation.  The mass spectrum corresponds to the mass scaling given by Eq. (\ref{pl}). The oscillations in the power law, called by authors fine structure, are obtained once the linear behavior is extracted; this can be accomplished without FMR. Thus in our context, the fine structure oscillations
for the super critical case are not calculated or related to the FMR method since the FMR method is used only in the subcritical case.}

Another feature (our main test for the FMR code) is the asymptotic structure of the spacetime at $\mathcal{J}^+$, particularly the Bondi mass and the news function. Figures 5 and 6 show them as a function of $\tau_B$ (the conditions are the same of Fig. 2), given by (\ref{tau}) but using the Bondi time and the accumulation Bondi time. As an extra test, Fig. 7 shows the global energy conservation with conditions as in Fig. 2.
{The periodicity in the Bondi mass 
   and the News function is as in P\"urrer et al. (Ref. [9]). 
   As expected by Eq. (12) the period of the Bondi mass has to be a half of the scalar field, and by Eq. (13) the period of the News function has to be the same as the scalar field. We pick up the 
   $r=0$ echoing for the scalar field 
   at null infinity for the Bondi mass and News function. Remarkably the energy 
   conservation test, a new result, is well behaved even in the threshold of the 
   black hole formation.}

\section{Discussion and conclusions}
We implemented the FMR method in the radial (compactified) coordinate for the characteristic formulation of General  Relativity. {Our final goal is to use the developed code in conjunction with other code which uses the ADM 3+1 formulation of General Relativity, in order to reach asymptotically an AdS spacetime. It is not possible in the present context to get an AdS boundary for a non-zero cosmological constant because the Bondi-Sachs coordinates are constructed for an asymptotically flat spacetime. Thus, in this work, we set the cosmological constant to zero.} As an important test we replicate the main features of the critical behavior in the collapse of a massless scalar field under spherical symmetry. We also replicate asymptotic quantities as the Bondi mass and News function. We obtain the energy conservation even in the extreme situation near the black hole formation, as an additional test and new result itself. 

{In Table 1 we display some indicators of the performance with and without FMR. Calculation of echoing without FMR, using $N_x=25,001$, shed a half of the cycle for $\lambda=0.14493045.$ This last calculation took 19 minutes on a  2.4 GHz Intel Core i5. Using $N_x=25,001$ distributed with our implementation of FMR takes 42 minutes. Thus, a better resolution of echoing is clear with FMR. Without FMR the resolution is not improved increasing $N_x$.}
\begin{table}[!ht]
\begin{tabular}{|l|c|c|c|c|c|}
\hline
\hline
{\sc FMR} & $\lambda$  & $N_x$  & Exec. time (min.) &cicle $(\Delta)$\\
\hline
{\sc Yes} &  $0.144930688869$   & 25,001 & 42  &   3/2 \\
\hline
{\sc No}  & 0.14493045\,\,\,\,\,\,\,\,\,\,\,\, & 25,001  & 19 & 1  \\
\hline
{\sc No}  & 0.14493045 \,\,\,\,\,\,\,\,\,\, &32,001 & 24 & 1 \\
\hline
\end{tabular}
\caption{Performance with and without FMR.}
\label{tabla}
\end{table}

{Pretorius and Lehner \cite{pl04} did an implementation of the AMR method in double null coordinates and they did a test for the particular case of the EKG system, far away from the critical behavior. We implemented the FMR in outgoing Bondi's coordinates and report as main test the subcritical critical behavior (echoing).}
Also, our results are in complete agreement with P\"urrer et al. \cite{petal05}. The comparison was focused mainly on the asymptotic behavior at null infinity. We use a different numerical method (the FMR) to get the echoing in the gravitational critical behavior at $\mathcal{J}^+$.}

{We conclude by stressing the following: At least in the characteristic
formulation, the implementation of the FMR method in 1-D (radial
coordinate) can be extended to non-spherical problems, because of the
spatial foliation structure $\mathcal{R}\times S^2$. In the 3-D case,
we can use the FMR orthogonal to the inflated cube technique to run
efficiently in parallel \cite{gbf07} for high angular definition.}

\section*{Acknowledgments}
The authors thank the financial support of Brazilian agencies CNPq and FAPERJ; also would like to thank Jennifer Rodriguez-Mueller for her valuable input to the paper.
{We thank the Referees because the presentation of our work indeed improves with their comments.}

\end{document}